\documentclass[manuscript]{aastex}

\slugcomment{To appear in ApJL}

\usepackage{graphicx}
\usepackage{natbib}
\shorttitle{Cold CO in the disk of EX Lup}
\shortauthors{K\'osp\'al et al.}

\begin{document}

\title{Cold CO gas in the disk of the young eruptive star
    EX Lup}


\author{\'A. K\'osp\'al\altaffilmark{1,2},
  P. \'Abrah\'am\altaffilmark{1}, T. Csengeri\altaffilmark{3},
  U. Gorti\altaffilmark{4,5}, Th. Henning\altaffilmark{2},
  A. Mo\'or\altaffilmark{1}, D. A. Semenov\altaffilmark{2},
  L. Sz\H{u}cs\altaffilmark{6}, R. G\"usten\altaffilmark{3}}

\altaffiltext{1}{Konkoly Observatory, Research
  Centre for Astronomy and Earth Sciences, Hungarian Academy of
  Sciences, P.O. Box 67, 1525 Budapest, Hungary}

\altaffiltext{2}{Max-Planck-Institut f\"ur Astronomie, K\"onigstuhl
   17, 69117 Heidelberg, Germany}

\altaffiltext{3}{Max-Planck-Institut f\"ur Radioastronomie, Auf dem
  H\"ugel 69, 53121 Bonn, Germany}

\altaffiltext{4}{SETI Institute, Mountain View, CA}

\altaffiltext{5}{NASA Ames Research Center, Moffett Field, CA}

\altaffiltext{6}{Max-Planck-Institut f\"ur Extraterrestrische
  Physik, 85741 Garching, Germany}


\begin{abstract}
EX~Lupi-type objects (EXors) form a sub-class of T Tauri stars,
defined by sudden sporadic flare-ups of 1-5 magnitudes at optical
wavelengths. These eruptions are attributed to enhanced mass accretion
from the circumstellar disk to the star, and may constitute important
events in shaping the structure of the inner disk and the forming
planetary system. Although disk properties must play a fundamental
role in driving the outbursts, they are surprisingly poorly known. In
order to characterize the dust and gas components of EXor disks, here
we report on observations of the $^{12}$CO J=3--2 and 4--3 lines, and
the $^{13}$CO 3--2 line in EX\,Lup, the prototype of the EXor class.
We reproduce the observed line fluxes and profiles with a line
radiative transfer model, and compare the obtained parameters with
corresponding ones of other T Tauri disks.
\end{abstract}



\section{Introduction}

EX~Lup is the prototype of EXors, one of the two main classes of young
eruptive stars. EXors are young low-mass T~Tauri stars exhibiting
repetitive outbursts due to a sudden increase of the accretion rate
\citep{herbig1977,herbig2008}. In 2008, EX~Lup exhibited its largest
outburst ever observed, triggering a series of multi-wavelength and
multi-epoch investigations of the accretion process. These led to the
discovery of the annealing of amorphous silicate particles to
crystalline grains during the outburst and their transportation to the
outer comet-forming zone \citep{abraham2009,juhasz2012}. Changes in
the molecular features of H$_2$O, OH, and H$_2$ indicate that the
outburst affected not only the surface mineralogy, but also the
chemistry in the inner disk \citep{banzatti2012}. Using high
resolution spectroscopic and spectroastrometric tools,
\citet{goto2011}, \citet{kospal2011}, and \citet{sicilia-aguilar2012}
concluded that the high accretion rate responsible for the outburst
mostly affected the innermost 0.4\,au region. They detected an
accretion-related wind and the motion of non-axisymmetric distribution
of material in the inner disk.

The emerging picture is broadly consistent with the magnetospheric
accretion scenario, in which the infalling material reaches the star
in hot spots. Indeed, X-ray and UV observations by \citet{grosso2010}
and by \citet{teets2012} indicate accretion shocks. Recently,
\citet{kospal2014} and \citet{sicilia-aguilar2015} studied the radial
velocity variations of optical absorption and emission lines, and
discussed different interpretation in terms of a possible brown dwarf
companion on a tight eccentric orbit around EX~Lup, or emission from
stable accretion columns in the system.

All the observations summarized above concentrated on the inner part
of the disk, within a few au of the star. However, the structure and
dynamics of the outer disk must play a significant role in driving the
outburst by replenishing the material in the inner disk after each
eruption. It is an open question whether the outer disks of EXors
differ in any way from the disks of normal T\,Tauri stars. A negative
answer could suggest that all low-mass stars undergo EXor phases
during their early evolution.

Very few information is available on the outer disk or on global disk
properties of the prototype, EX\,Lup. The emission of the cold dust
component was detected at infrared wavelengths
\citep{grasvelazquez2005,sipos2009}, and at 870\,$\mu$m with
APEX/LABOCA \citep{juhasz2012}. Modeling the spectral energy
distribution (SED) based on these data, \citealt{sipos2009} and
\citet{juhasz2012} deduced a modestly flared disk geometry. In the
lack of spatially resolved infrared or millimeter data, the outer disk
radius of EX\,Lup is unknown. Much less is known about the gas
component. Warm carbon monoxide (CO) gas was detected in the
fundamental lines at 4.5$\,\mu$m and in the overtone band at
2.3$\,\mu$m \citep{aspin2010,goto2011,kospal2011}. The profiles of the
fundamental lines could be well fitted with an inner disk inclination
angle between 40$^{\circ}$ and 50$^{\circ}$
\citep{goto2011}. Concerning the cold gas in the disk, the only
available millimeter CO observation reported in the literature, a
$^{12}$CO(3--2) line targeted by \citet{vankempen2007} using JCMT, was
a non-detection.

In order to study the abundance, distribution, and kinematics of cold
molecular gas in EX\,Lup's disk, we obtained new submillimeter CO line
observations with APEX/FLASH$^+$. In this Letter, we present the data
and analyze the line profiles, line intensities, measure the CO mass,
optical depth, reproduce the results with a chemical and radiative
transfer model, and confront the results with those obtained for
typical T\,Tauri-type stars.


\section{Observations}

We used the FLASH$^+$ receiver \citep{klein2012} at the APEX
telescope~{\citep{gusten2006}} to measure the $^{12}$CO(3--2),
$^{13}$CO(3--2), and $^{12}$CO(4--3) lines towards EX Lup on 2015
March 30 -- April 1. The lower frequency channel was tuned to
344.2\,GHz in USB to cover the $^{13}$CO(3--2) at 330.588\,GHz, and
the CO(3--2) at 345.796\,GHz, respectively. The higher frequency
channel was tuned to the $^{12}$CO(4--3) line at 461.041\,GHz in
USB. We used the XFFTS backends providing a nominal 38.15\,kHz
spectral resolution. Observations were carried out in position
switching mode with a relative reference position 100$''$ away.

The spectra have been averaged and a first order baseline has been
removed. The noise level on a T$_A^*$ scale in 1 km/s channels is
5.2\,mK (0.21\,Jy) for $^{12}$CO(3--2), 11.5\,mK (0.55\,Jy) for
$^{12}$CO(4--3), and 7.7\,mK (0.32\,Jy) for $^{13}$CO(3--2). We used a
Jy-to-K conversion of 41\,Jy/K for the 3--2 lines and 48\,Jy/K for the
4--3 line to convert from T$_A^*$ scale to flux scale. The telescope's
beam is $19$\farcs$2$, and $15$\farcs$3$ at the corresponding
frequencies.


\section{Results and Analysis}
\label{sec:results}

\begin{figure}
\includegraphics[angle=90,scale=.68]{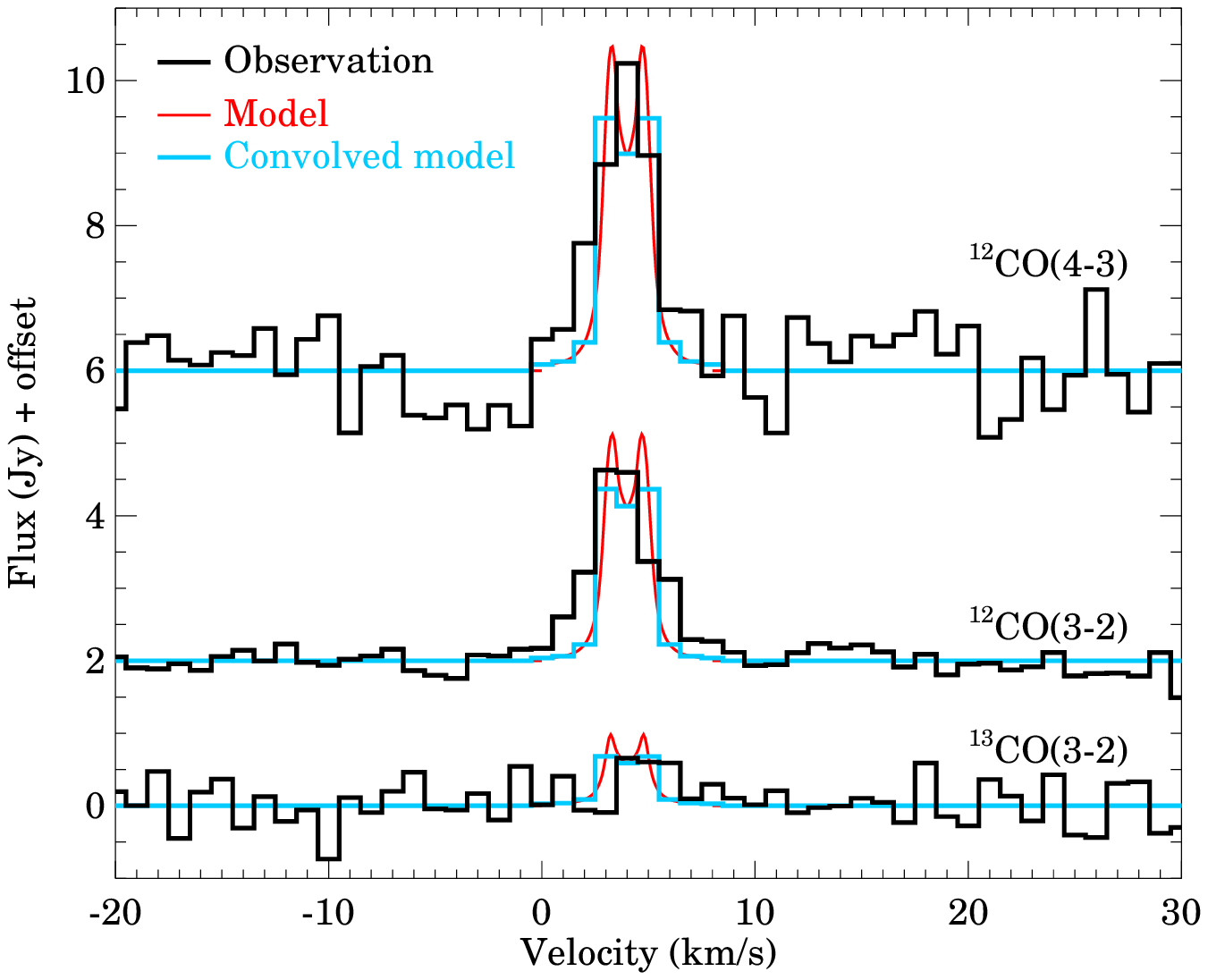}
\caption{CO line profiles of EX Lup observed with APEX/FLASH$^+$
  (black histograms). The red curves are from our line radiative
  transfer model (for details, see Sec.~\ref{sec:model}). The blue
  histograms are the models convolved to the same spectral resolution
  as the observations.
\label{fig:exlupco}}
\end{figure}

Our CO spectra are plotted in Fig.~\ref{fig:exlupco}. The $^{12}$CO
lines are securely detected at 12$\sigma$ and 7.7$\sigma$ levels for
the 3--2 and 4--3 lines, respectively. There is a marginal detection
of the $^{13}$CO(3--2) line at a level of 2.6$\sigma$, at the same
radial velocity as the other CO lines. The observed lines are
single-peaked. The peak fluxes, total line intensities, line widths
(FWHM of fitted Gaussians), and line positions (centers of the fitted
Gaussians) are presented in Tab.~\ref{tab:results}.

The ratio of the 3--2 lines of the two different CO isotopologues can
be used to calculate the optical depths of the lines. If we denote the
optical depths of $^{12}$CO and $^{13}$CO by $\tau_{12}$ and
$\tau_{13}$, respectively, then the ratio of the $^{12}$CO to the
$^{13}$CO line is approximately ($1 - e^{-\tau_{12}}$)/($1 -
e^{-\tau_{13}}$). Assuming that the optical depths of the different
isotopologues follow the same proportions as the abundance ratios
typical of local interstellar matter \citep{wilson1999}, $\tau_{12}$ =
69$\times \tau_{13}$. Using these numbers, we obtained $\tau_{12}$ $\approx$
20 and $\tau_{13}$ $\approx$ 0.3. Even considering the uncertainties in this
evaluation, it is very likely that the $^{12}$CO lines are optically
thick, while the $^{13}$CO line is optically thin.

The temperature of the CO gas can be estimated from the ratio of the
optically thick $^{12}$CO(4--3) and $^{12}$CO(3--2) lines. In the
Rayleigh--Jeans approximation, the ratio is expected to be the ratio
of the squares of the line frequencies, i.e., about 1.78. Instead,
using the total flux values from Tab.~\ref{tab:results}, we obtain
1.25$\pm$0.15, significantly different from the Rayleigh--Jeans
value. The low value suggests that the temperature is very low and
that the Rayleigh--Jeans approximation is not valid. Indeed, using the
Planck function, this line ratio corresponds to 10$^{+4}_{-2}$\,K.

From the optically thin $^{13}$CO line, using the canonical 10$^{-4}$
CO-to-H$_2$ abundance ratio, we estimate a total disk mass of
2.3$\times$10$^{-4}$\,M$_{\odot}$. If the abundance ratio is lower
than 10$^{-4}$, as will be discussed in Sect.~\ref{sec:codepletion},
then this disk mass should be considered as a lower limit.


\section{Chemical and Radiative Transfer Modeling}
\label{sec:model}

In order to reproduce the observed line profiles and line fluxes, we
made a detailed chemical and radiative transfer model of the EX\,Lup
disk. As a base, we used the model of \citet{sipos2009} to fit the
quiescent spectral energy distribution (SED), providing radial and
vertical dust density and dust temperature distributions. We used this
physical model and performed a detailed simulation of the disk
chemistry similarly to that described in \citet{gorti2011}. The CO
abundance, densities, and temperatures were calculated without
assuming local thermodynamical equilibrium (LTE), and included
radiative pumping, dust background radiation, and spontaneous emission
and collisions. In parallel, we verified the results with the
state-of-the-art chemical code ALCHEMIC \citep{semenov2010}, and found
similar results. The obtained gas-phase $^{12}$CO fractional
abundances were used as input for our line transport calculations,
performed with the {\sc radmc-3d}
code\footnote{http://www.ita.uni-heidelberg.de/\~{}dullemond/software/radmc-3d/}.
We assumed a homogeneous gas-to-dust ratio of 100 and a Keplerian
velocity field with a microturbulent velocity of 0.1 km/s, taken to be
constant throughout the disk. The abundances of $^{13}$CO and
C$^{18}$O were calculated using a constant $^{12}$CO/$^{13}$CO ratio
of 69 and $^{12}$CO/C$^{18}$O ratio of 560 \citep{wilson1999}.  The
obtained densities and temperatures imply that LTE holds everywhere in
the disk, thus we adopted LTE for modeling the line profiles. The gas
temperature was assumed to be equal with the dust temperature. The
line emission was calculated in 8\,km/s wide windows centered on the
rest frame wavelengths of the J=3--2 and 4--3 transitions of $^{12}$CO
and $^{13}$CO.

In Fig.~\ref{fig:exlupco} we overplotted the resulting model
spectra. In order to match the brightness of the observed lines, the
gas temperature in our model had to be scaled down by a factor of
0.8. With this small modification, all measured line fluxes and line
ratios are well reproduced by our model, indicating that the disk
parameters responsible for the line width were reasonable
estimates. This result also confirms that the inclination of
40$^{\circ}$ used in the radiative transfer calculations is good
approximation. The double-peaked profiles characteristic for a disk in
Keplerian orbit are not observed probably due to the limited spectral
resolution of the data. We emphasize that while our modeling provides
a possible solution to reproduce the observed lines, it was not a fit
to the lines. The exploration of the full parameter space needed to
prove that this is a unique solution is beyond the scope of this
Letter. For this reason, in the following, we base our discussion
on our measured CO line fluxes.


\section{Discussion}

\subsection{Disk mass from dust continuum data}

\citet{sipos2009} assumed small grains and fitted the SED with a total
disk mass of 0.025\,M$_{\odot}$. \citet{sicilia-aguilar2015} could
obtain a reasonably well fit to the SED with a total mass of
1--3$\times$10$^{-3}$\,M$_{\odot}$ by using grains between 0.1 and
100$\,\mu$m with collisional size distribution. In these studies, when
converting the dust mass to total mass, a gas-to-dust mass ratio of
100 was assumed. While this is typical for the interstellar medium
(ISM), circumstellar disks often have lower gas-to-dust mass
ratios. Recently, \citet{williams2014} measured values between 43 and
2 for T\,Tauri stars in Taurus, based on Submillimter Array (SMA)
observations of the 1.3\,mm continuum and $^{13}$CO(2--1) and
C$^{18}$O(2--1) lines.
Considering these uncertainties, the total disk mass of EX\,Lup from
dust continuum data may be as high as 0.025\,M$_{\odot}$ (using small
grains and gas-to-dust ratio of 100) or as low as a few times
10$^{-4}$\,M$_{\odot}$ (using a grain size distribution and a
gas-to-dust ratio of 10). While the lower value is roughly consistent
with the disk mass obtained from our $^{13}$CO data, the higher value
is about 100 times higher, hinting at significant CO depletion in the
disk of EX\,Lup.

\subsection{CO depletion in EX\,Lup}
\label{sec:codepletion}

\begin{figure}
\includegraphics[angle=90,scale=.59]{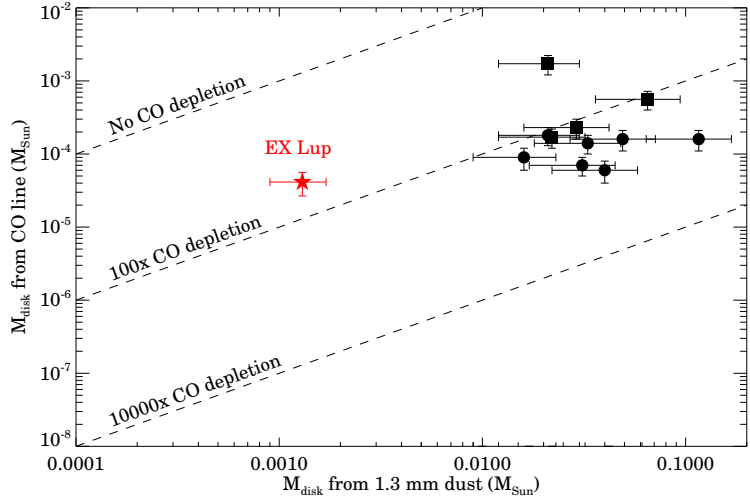}
\caption{Disk masses estimated from CO observations against disk
  masses from dust continuum data (based on Fig.~10 of
  \citealt{thi2001}). The black circles indicate T\,Tauri stars, the
  black squares mark Herbig Ae/Be stars, while EX\,Lup is plotted with
  a red asterisk.\label{fig:thi}}
\end{figure}

In order to find out whether the disk properties derived from our CO
observations of EX\,Lup are in any way special, we need to compare it
to normal, non-eruptive young stars. \citet{thi2001} observed the
$^{13}$CO(3--2) line in eight T\,Tauri-type stars and in seven Herbig
Ae stars in the Taurus-Auriga cloud, calculated the total gas masses,
and compared them with total masses inferred from 1.3\,mm dust
continuum measurements. They found that ``masses derived from CO are
generally 10-200 times lower than those found from the millimeter
continuum''. In Fig.~\ref{fig:thi} we reproduced Fig.~10 from
\citet{thi2001}, and overplotted EX\,Lup, using the same equations and
assumptions to calculate the disk masses as was used in that paper (we
estimated a range of 1.3\,mm fluxes for EX\,Lup by extrapolating from
the 870$\,\mu$m flux presented in \citealt{juhasz2012} using
$\beta$=0...1.7 as the spectral index of the dust opacity). Compared
to the sample of \citet{thi2001}, EX\,Lup has a remarkably small disk
mass and modest CO depletion. This result means that the total
  disk mass might be a factor of 10...100 higher than the value
  calculated in Sect.~\ref{sec:results}.

\citet{thi2001} listed two possible reasons for the CO depletion: (1)
freeze-out in the coldest, mid-plane regions of the disk (indeed,
\citet{reboussin2015} found that the canonical gas-phase abundance of
CO compared to H$_2$ of 10$^{-4}$ is only reached above about 30 --
35\,K, below this temperature, the ratio is much smaller because CO
depletion due to freeze-out is very efficient); (2) photodissociation
by stellar or interstellar UV radiation in the disk surface. These two
effects may operate in EX\,Lup as well. Firstly, according to our
radiative transfer model, the dust temperature in the disk midplane in
the outer disk regions is indeed below 15\,K, therefore CO is expected
to freeze out. Secondly, the two M-type stars in the sample of
\citet{thi2001} show similar CO depletion factor (about 100),
indicating similar radiation fields.

The fact that the CO depletion in EX\,Lup is less than those observed
in normal T\,Tauri and most of those in Herbig Ae stars as well has
interesting implications. EX\,Lup exhibited a large outburst in 2008,
when both its optical and X-ray brightness increased by orders of
magnitude \citep{kospal2008,teets2012}. According to our modeling,
both the midplane and the surface temperature increased in the disk
during outburst \citep{abraham2009}. On the one hand, the increased
temperature might have evaporated CO ice and increased the abundance
of gas-phase CO, as predicted by \citet{vorobyov2013b}. On the other
hand, the increased UV flux during outburst might have
photodissociated a significant amount of CO gas. It seems that in the
case of EX\,Lup, the latter effect was more dominant. Our low-J CO
observations trace the bulk of the cold gas, therefore the outburst
affected the outer parts of the disk, and not only its innermost part
(where \citealt{banzatti2015} found a depletion of hot CO after the
outburst, interpreted as depletion of material accumulated around and
within the corotation radius at 0.02--0.3\,au). Our observations
support the conclusions of \citet{vorobyov2013b}, whose numerical
simulations showed that the chemical signatures of luminosity bursts
in the gas-phase CO abundance may linger for several thousand years.

\subsection{The mechanism of EXor outbursts?}

\citet{baobab} observed four EXors in the 1.3\,mm continuum with the
SMA. They detected two targets, one with a relatively high dust mass
(NY\,Ori, 9$\times$10$^{-4}$\,M$_{\odot}$), and one with much lower
(V1118\,Ori, 6$\times$10$^{-5}$\,M$_{\odot}$). The other two targets
were undetected, with 3$\,\sigma$ upper limits of
6$\times$10$^{-5}$\,M$_{\odot}$ for V1143\,Ori and
6$\times$10$^{-6}$\,M$_{\odot}$ for VY\,Tau. Compared to these, the
dust disk mass of EX\,Lup is in the range of the least massive EXor
disks detected. Therefore, the prototype of the class fits into the
trend that EXors disks are less massive than FUor disks. A possible
cause for the low disk masses is binarity: there is a trend that stars
with companions within 100\,au generally have lower disk masses than
single stars or wider binaries \citep{osterloh1995,
  andrews2005}. Indeed, many EXors are binaries (e.g., VY\,Tau,
V1118\,Ori, or XZ\,Tau, see \citealt{leinert1993},
\citealt{reipurth2007}, \citealt{hartigan2003}). EX\,Lup have been
searched for companions with different methods without conclusive
results (see, e.g., \citealt{kospal2014} and references therein).

The low EXor disk masses raise the question what mechanism causes the
eruption of these disks and what reservoir can replenish the material
in the inner disk after it lands on the star. The low disk masses
virtually exclude mechanisms related to gravitational instability. A
promising new idea was presented by \citet{dangelo2012}, who proposed
that accretion onto a strongly magnetic protostar is inherently
episodic if the disk is truncated close to the corotation
radius. \citet{dangelo2012} demonstrated that this mechanism may work
for EX\,Lup.


\acknowledgments

This work was supported by the Momentum grant of the MTA CSFK
Lend\"ulet Disk Research Group, and the Hungarian Research Fund OTKA
grant K101393. A.~M.~acknowledges support from the Bolyai Research
Fellowship of the Hungarian Academy of Sciences.

{\it Facilities:} \facility{APEX}.



\begin{deluxetable}{cccccc}
\tabletypesize{\scriptsize}
\tablecaption{CO Observations of EX Lup.\label{tab:results}}
\tablewidth{0pt}
\tablehead{
\colhead{Line} & \colhead{Frequency} & \colhead{Peak} & \colhead{Flux}       & \colhead{Width}  & Position  \\
               & \colhead{(GHz)}     & \colhead{(Jy)} & \colhead{(Jy\,km/s)} & \colhead{(km/s)} & \colhead{(km/s)}}
\startdata
$^{12}$CO(3--2) & 345.796   & 2.63 $\pm$ 0.21 & 9.74 $\pm$ 0.41  & 3.6 $\pm$ 0.2 & 3.6 $\pm$ 0.2 \\
$^{12}$CO(4--3) & 461.041   & 4.24 $\pm$ 0.55 & 12.15 $\pm$ 1.34 & 2.6 $\pm$ 0.3 & 3.9 $\pm$ 0.2 \\
$^{13}$CO(3--2) & 330.588   & 0.82 $\pm$ 0.32 & 2.09 $\pm$ 0.74  & 2.7 $\pm$ 1.1 & 5.1 $\pm$ 0.5 
\enddata
\end{deluxetable}

\end{document}